\documentclass[preprint,12pt]{elsarticle}

\usepackage{amssymb}
\usepackage{multirow}

\journal{Nuclear Instruments and Methods in Physics Research Section A}

\begin{document}

\begin{frontmatter}



\title{Development of High Intensity Laser-Electron Photon Beams up to 2.9~GeV
at the SPring-8 LEPS Beamline}


\author[label1,label2]{N.~Muramatsu}
\author[label2]{Y.~Kon}
\author[label3]{S.~Dat\'{e}}
\author[label3]{Y.~Ohashi}
\author[label4]{H.~Akimune}
\author[label2]{J.Y.~Chen}
\author[label2]{M.~Fujiwara}
\author[label5]{S.~Hasegawa}
\author[label2]{T.~Hotta}
\author[label1]{T.~Ishikawa}
\author[label6]{T.~Iwata}
\author[label7]{Y.~Kato}
\author[label2]{H.~Kohri}
\author[label8]{T.~Matsumura}
\author[label9]{T.~Mibe}
\author[label6]{Y.~Miyachi}
\author[label2]{Y.~Morino}
\author[label2]{T.~Nakano}
\author[label2]{Y.~Nakatsugawa}
\author[label3]{H.~Ohkuma}
\author[label2]{T.~Ohta}
\author[label2]{M.~Oka}
\author[label2]{T.~Sawada}
\author[label10]{A.~Wakai}
\author[label11]{K.~Yonehara}
\author[label2]{C.J.~Yoon}
\author[label2]{T.~Yorita}
\author[label2]{M.~Yosoi}

\address[label1]{\footnotesize Research Center for Electron Photon Science, Tohoku University, Sendai, Miyagi 982-0826, Japan}
\address[label2]{\footnotesize Research Center for Nuclear Physics, Osaka University, Ibaraki, Osaka 567-0047, Japan}
\address[label3]{\footnotesize Japan Synchrotron Radiation Research Institute, Sayo, Hyogo 679-5143, Japan}
\address[label4]{\footnotesize Department of Physics, Konan University, Kobe, Hyogo 658-8501, Japan}
\address[label5]{\footnotesize J-PARC Center, Japan Atomic Energy Agency, Tokai-mura, Ibaraki 319-1195, Japan}
\address[label6]{\footnotesize Department of Physics, Yamagata University, Yamagata 990-8560, Japan}
\address[label7]{\footnotesize Department of Physics, Nagoya University, Nagoya, Aichi 464-8602, Japan}
\address[label8]{\footnotesize National Defence Academy in Japan, Yokosuka, Kanagawa 239-8686, Japan}
\address[label9]{\footnotesize High Energy Accelerator Research Organization, KEK, Tsukuba, Ibaraki 305-0801, Japan}
\address[label10]{\footnotesize Akita Research Institute of Brain and Blood Vessels, Akita, 010-0874, Japan}
\address[label11]{\footnotesize Illinois Institute of Technology, Chicago, Illinois 60616, USA}

\begin{abstract}

  A laser-Compton backscattering beam, which we call a `Laser-Electron Photon' beam,
  was upgraded at the LEPS beamline of SPring-8. We accomplished the gains in 
  backscattered photon beam intensities by factors of 1.5--1.8 with the injection 
  of two adjacent laser beams or a higher power laser beam into the storage ring.
  The maximum energy of the photon beam was also extended from 2.4~GeV to 2.9~GeV 
  with deep-ultraviolet lasers. The upgraded beams have been utilized for hadron
  photoproduction experiments at the LEPS beamline. Based on the developed methods,
  we plan the simultaneous injection of four high power laser beams at the LEPS2
  beamline, which has been newly constructed at SPring-8. As a simulation result,
  we expect an order of magnitude higher intensities close to 10$^7$~sec$^{-1}$ 
  and 10$^6$~sec$^{-1}$ for tagged photons up to 2.4~GeV and 2.9~GeV, respectively.

\end{abstract}


\end{frontmatter}

% \linenumbers


\section{Introduction} \label{sec:intro}

    Backward Compton scattering of laser light from a high energy electron beam \cite{milburn,arutyunyan} 
  is a popular technique to produce $\gamma$-ray beams in a wide energy range. We refer to such a $\gamma$-ray 
  beam as a `Laser-Electron Photon' (LEP) beam. In the last decade, several facilities to produce the LEP 
  beam have been successfully operated at high-current electron storage rings, where the coexistence with 
  synchrotron radiation light sources is reasonably achieved. Energies of backscattered photons can be 
  magnified nearly up to the same order of magnitude as the electron beam energy by injecting ultraviolet 
  (UV) laser light into a multi-GeV storage ring. The LEP beams generated in this way at the energies from 
  $\sim$100~MeV to a few GeV have been used in the experiments for hadron and nuclear physics, as summarized 
  in Ref.~\cite{muramatsu,dangelo}. The number of facilities with low energy beams up to a few tens MeV using 
  an infrared (IR) laser or a low energy storage ring is also increasing for nuclear physics and astrophysics 
  \cite{ohgaki,aoki,kawase,guo,ahn}. 

    The LEP beam possesses several advantages for the physics experiments as follows: Firstly, 
  a photon beam is clean without hadron contaminations. Although pair-created electrons and positrons 
  are unavoidably contaminated, they are removed by a sweeping magnetic field, a charge veto counter, 
  and a ${\rm \check{C}}$erenkov counter. Secondly, a spread of the backscattered photon directions
  is roughly equivalent to a polar angle of 1/$\gamma$ rad, where $\gamma$ represents the Lorentz factor 
  for multi-GeV electrons. A collimated beam can be provided to an experimental site. Thirdly, an energy 
  spectrum of the LEP beam is rather flat below the Compton edge, compared with that of a bremsstrahlung 
  photon beam. Backgrounds due to low energy photon reactions, which are usually untagged, are relatively
  suppressed in the case of the LEP beam. Finally, a LEP beam possesses nearly 100\% polarization at 
  the maximum energy because laser polarization is retained at a hard head-on collision of a laser photon 
  and an electron. A polarization vector of the LEP beam is easily controllable by handling the laser 
  polarization, which can be linear, circular, or elliptical.

\begin{figure}[htbp]
   \centering
   \includegraphics[width=7cm]{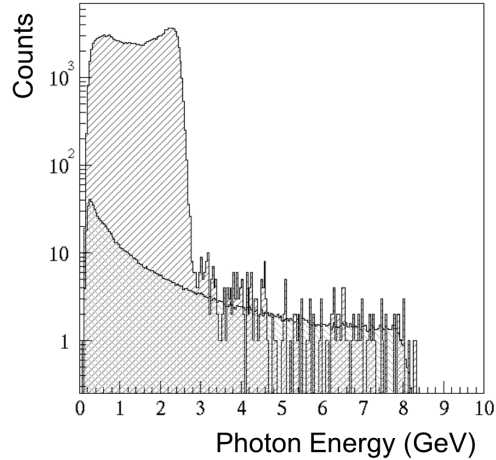}
   \caption{A LEP energy spectrum with the injection of a multiline UV Ar laser light (the hatched 
            area by parallel lines). The photon energies were measured by a calorimeter made of PWO 
            crystals \cite{matsumura}, which were installed on the LEP beam path. A spectrum for 
            bremsstrahlung photons was also measured without laser injection (the hatched area by 
            crossed lines).}
   \label{fig:espect}
  \end{figure}

    So far, the highest LEP energy region has been covered by the LEPS beamline at SPring-8 \cite{spring8}, 
  which stores 7.975~GeV electrons with a current of 100~mA. From a kinematical calculation, the maximum 
  energy of the LEP beam ($k_{max}$) is given by:
  \begin{equation}
      k_{max} = \frac{(E_e + P_e c) k_{laser}}{E_e - P_e c + 2 k_{laser}}
         \simeq \frac{4 {E_e}^2 k_{laser}}{{m_e}^2 c^4 + 4 E_e k_{laser}} ,  \label{eqn:maxene}
  \end{equation}
  where $E_e$, $P_e$, and $m_e$ represent the energy, momentum, and mass of an electron in the storage 
  ring, respectively, and $k_{laser}$ denotes the energy of laser light. At an early stage of the LEPS
  experiments, a multiline UV argon-gas laser (Ar laser) with output wavelengths around 351~nm was 
  operated \cite{leps-init} as done widely in other facilities. A LEP energy spectrum with the Ar laser 
  has been observed as shown in Fig.~\ref{fig:espect}, indicating the maximum energy of 2.4~GeV in 
  accordance with Eq.~\ref{eqn:maxene}. In Fig.~\ref{fig:espect}, a spectrum for bremsstrahlung photons,
  arising from the electron deceleration by a residual gas inside the storage ring, is overlaid by 
  normalizing event entries in the energy range above 2.4~GeV. The comparison of the two spectra 
  suggests that the intensity of a LEP beam is more than two orders of magnitude higher than that of
  a bremsstrahlung photon beam. The generated LEP beam was transported to a fixed target made of liquid 
  hydrogen or deuterium. The liquid target was sometimes replaced to a solid nuclear target. Structures 
  and interactions of hadrons, particularly including a strange quark, have been studied by measuring 
  differential cross sections and polarization observables in the photoproduction from those targets, 
  as summarized in Ref.~\cite{muramatsu}.

    Although the LEP beam production by using the Ar laser has been established with the successful 
  results, further improvements concerning the beam intensity and the maximum energy are desired to 
  advance hadron and nuclear physics. Firstly, an interaction rate of a photon beam is lower than 
  that of a hadron beam, so that a higher intensity LEP beam is needed for precise and systematic 
  measurements. For this purpose, a method to inject multiple laser beams has been developed 
  as discussed later. In addition, a UV output power is increasing thanks to the progress in 
  the technologies of all-solid lasers. Secondly, the extension of the maximum LEP energy opens 
  new fields of the photoproduction experiments. For example, it becomes feasible to study 
  the photoproduction of heavy particles, like the K$^{*}$(892) production in the t-channel.
  A recoilless condition of $\rho$ and $\omega$ mesons, which are produced inside a nucleus, is 
  also achieved at $E_\gamma$ $=$ 2.75~GeV, possibly making nuclear medium modification more 
  efficient. High power deep-UV (DUV) lasers with a shorter wavelength are now available, and
  the production of a high energy LEP beam is hence realistic as recognized from Eq.~\ref{eqn:maxene}.

    The following part of the present article describes the results on the intensity and 
  energy upgrades at the LEPS beamline together with the studies for a near-future project. 
  In Section~\ref{sec:setup}, experimental setups for the LEP beam production and upgrades 
  are explained in addition to the methods of intensity and energy measurements.
  In Section~\ref{sec:achievements}, we discuss the results obtained by introducing the new 
  solid-state laser, the DUV lasers, and the simultaneous two-laser injection method. 
  In Section~\ref{sec:developments}, we argue new attempts to increase a LEP intensity and 
  simulation studies for a new beamline, whose construction has just finished for carrying
  out next-generation experiments at SPring-8. Finally, a brief summary follows 
  in Section~\ref{sec:summary}.

\section{Experimental Setup} \label{sec:setup}

 \subsection{UV and DUV Lasers} \label{sec:laser}

    Table~\ref{tab:uvlasers} shows the properties of the UV lasers which have been used at 
  the LEPS experiments. The Ar laser (`Innova Sabre') was operated from the beginning of 
  the experiments, which started in 1999. It provides an output power of 7~W with three 
  major contributions from the wavelengths of 333.6--335.8~nm, 351.1~nm, and 363.8~nm. 
  These contributions produce the Compton edges at 2.49~GeV, 2.40~GeV, and 2.34~GeV, 
  respectively, with the strengths similar to each other. For the discharge inside 
  an argon-gas tube, the whole system typically required 10~kW electric power consumption 
  with a three-phase AC input of 480~V. Since 2006, the `Innova Sabre' has been replaced 
  to the solid-state laser `Paladin' with a UV output power of 8 W. Its power consumption 
  is reduced to 350~W with a 100~V / 200~V AC power supply. This low power consumption is 
  achieved by introducing a combination of a seed laser diode and a nonlinear crystal, 
  which triples a seed laser frequency by third harmonic generation (THG). The output 
  wavelength of the `Paladin' is only 355~nm, and the Compton edge of the LEP beam with
  this laser has become 2.39~GeV. In Sections~\ref{sec:modelock} and \ref{sec:2laser}, 
  the `Paladin' with the 8~W UV power (8~W `Paladin') was used for further development of 
  LEP beams. An upgraded version of this laser with an output power of 16~W (16~W `Paladin')
  was also tested as described in Section~\ref{sec:newtest1}.
  \begin{table}[htbp]
  \caption{Properties of the UV lasers operated at the LEPS beamline. The `$1/e^2$ diameter' 
           is defined for the circular region where the beam density is higher than the $1/e^2$ 
           of the peak value.}
  \centering
  \begin{tabular}{lccc}
  \hline
   \multirow{3}{*}{Laser name}         &    Innova Sabre     &     Paladin             &      Paladin            \\
                                       &      DBW25/7        &     355-8000            &      355-16000          \\
                                       &   (Coherent Inc.)   &  (Coherent Inc.)        &   (Coherent Inc.)       \\
   \hline
   \multirow{2}{*}{Wavelength}         &    multiline UV     & \multirow{2}{*}{355 nm} & \multirow{2}{*}{355 nm} \\
                                       &   333.6--363.8 nm   &                         &                         \\
   \multirow{2}{*}{Emission frequency} & \multirow{2}{*}{CW} &      80 MHz             &       80 MHz            \\
                                       &                     &    (quasi-CW)           &     (quasi-CW)          \\
   UV output power                     &       7 W           &       8 W               &       16 W              \\
   $1/e^2$ diameter                    &      1.7 mm         &      1.0 mm             &      1.35 mm            \\
   Divergence                          &     0.31 mrad       &    0.55 mrad            &     0.55 mrad           \\
   Power consumption                   &      10 kW          &      350 W              &       350 W             \\
  \hline
  \end{tabular}
  \label{tab:uvlasers}
  \end{table}

    The DUV lasers which are listed in Table~\ref{tab:duvlasers} have been introduced 
  for the extension of the maximum energy at the LEPS beamline. All the lasers supplies 
  a continuous-wave (CW) beam with a shorter wavelength of 257~nm or 266~nm, which is 
  achieved by doubling the frequency of a seed green laser (second harmonic generation, 
  SHG) at a Barium Borate (BBO) crystal. After confirming a LEP energy spectrum by using 
  `DeltaTrain' as described in Section~\ref{sec:duvinj}, the collection of large physics 
  data have been done with `Innova Sabre Moto FreD' and `Frequad-HP', whose output powers 
  are both 1~W. A seed laser of the `Innova Sabre Moto FreD' is the argon-gas tube system,
  which has been operated as the resonator of the `Innova Sabre', and it requires huge
  power consumption. Therefore, the method of two-laser injection has been developed by 
  using the `Frequad-HP', which is an all-solid DUV laser. The LEP beam intensities with 
  these lasers and the two-laser injection are discussed in Sections~\ref{sec:duvinj} 
  and \ref{sec:2laser}, respectively.
  \begin{table}[htbp]
  \caption{Properties of the DUV lasers operated at the LEPS beamline.}
  \centering
  \begin{tabular}{lccc}
  \hline
   \multirow{3}{*}{Laser name} &       DeltaTrain       &    Innova Sabre     &   Frequad-HP    \\
                               &    (Spectra-Physics    &      Moto FreD      &  (Oxide Corp.)  \\
                               &          Inc.)         &   (Coherent Inc.)   &                 \\
   \hline
   Wavelength                  &        266 nm          &     257.2 nm        &     266 nm      \\
   Emission frequency          &          CW            &       CW            &      CW         \\
   UV output power             &        0.7 W           &       1 W           &      1 W        \\
   $1/e^2$ diameter            &     0.5--0.6 mm        &   0.6--0.9 mm       &     3.0 mm      \\
   Divergence                  &     0.6--0.9 mrad      &   0.5--0.85 mrad    &     0.4 mrad    \\
   Power consumption           &        370 W           &      10 kW          &     300 W       \\
  \hline
  \end{tabular}
  \label{tab:duvlasers}
  \end{table}

 \subsection{Laser Beam Focus} \label{sec:focus}

\begin{figure}[htbp]
   \centering
   \includegraphics[width=14cm]{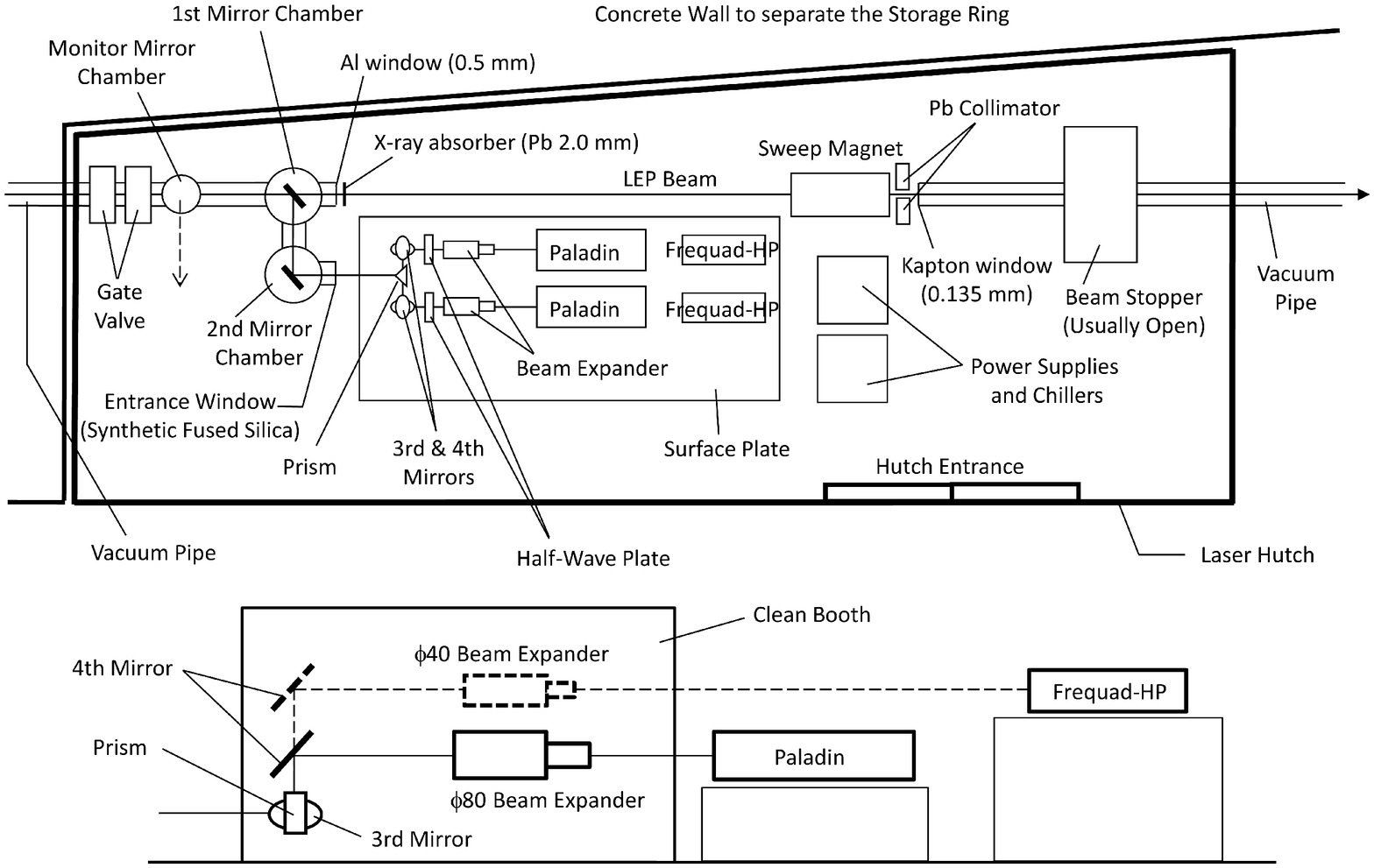}
   \caption{A plan view of the experimental setup inside the laser hutch (upper part). Lasers, 
            optical control devices, and vacuum chambers are installed to inject laser beams 
            into the electron storage ring, which is located in the left-hand side. A side 
            view of the injection system on a surface plate is also shown in the lower part.}
   \label{fig:hutch}
  \end{figure}

\begin{figure}[htbp]
   \centering
   \includegraphics[width=7cm]{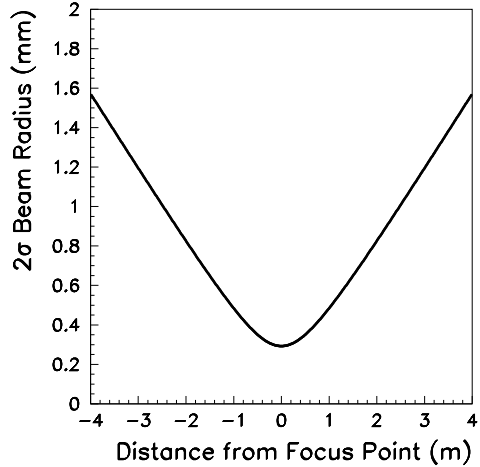}
   \caption{The 2$\sigma$ ($1/e^2$) radius of 355~nm laser light from the 8 W `Paladin' at the 
            straight section of LEPS beamline. An ideal beam propagation from a 37~m-upstream 
            expander is assumed in the calculation based on Eq.~\ref{eqn:propagation}. At the 
            expander output, a laser beam diameter is expanded from 1~mm to 28.6~mm.}
   \label{fig:355waist}
  \end{figure}

    We developed several methods to improve the LEP beam properties at the LEPS beamline 
  (BL33LEP) of SPring-8. Figure~\ref{fig:hutch} shows a schematic view of the experimental 
  setup to inject laser beams into the electron storage ring. The whole injection system 
  was set up inside the laser hutch, which is an interlocked radiation shield room beside 
  a thick concrete wall of the ring tunnel. Lasers and injection optics were placed on 
  a large surface plate with coverage by a clean booth. Since the injected laser light 
  collides with a narrow electron beam at the straight section of the storage ring after 
  traveling a distance of 37~m, it is necessary to introduce a beam expander, which once 
  enlarges a laser diameter for making a focus (or a beam waist) at the collision point. 
  For an ideal Gaussian beam, a $1/e^2$ or 2$\sigma$ radius ($r$~[mm]) at an arbitrary 
  distance from the focus point ($z$~[m]) is uniquely determined by a desired waist radius 
  ($w_0$~[mm]) as described with the following equation:
  \begin{equation}
      r = w_0 \times \sqrt{1+(\lambda \times |z| / (\pi \times w_0^2 \times 10^3))^2} , \label{eqn:propagation}
  \end{equation}
  where $\lambda$ represents a wavelength of the laser in nm. In our case with the 8~W 
  `Paladin', the $1/e^2$ radius around the beam waist is calculated as shown in
  Fig.~\ref{fig:355waist}. Because the average standard deviations ($\sigma$) of the electron 
  beam at the straight section are 295~$\mu$m and 12~$\mu$m in the horizontal and vertical 
  directions, respectively, the $1/e^2$ radius desired for the laser beam waist is about 
  300~$\mu$m from an average of the two electron beam sizes. Consequently, the magnification 
  factors to determine the output diameters from the beam expanders were set to 28.6 and 13.3 
  for the laser light from the 8~W `Paladin' and the `Frequad-HP', respectively.

    A structure of the beam expander is the same as a Galilean telescope, which increases 
  an input laser diameter at a set of small concave lenses and outputs a convergent beam 
  from a large convex lens. These optical lenses were made of high grade synthetic fused 
  silica, which minimizes chemical impurities in order to reduce the generation of color 
  centers by a high power laser. In contrast, a modest amount of hydroxyl (OH) concentrations 
  at $\sim$1000~ppm is accepted for UV radiation resistance \cite{hydroxyl}. Beam expanders 
  for UV and DUV lasers were separately prepared to ensure high transmission rates with 
  individually optimized anti-reflection (AR) coating on the lens surfaces. The convex 
  lens diameter at the expander output was set to 40~mm ($\phi$40) for the DUV laser, 
  while this diameter was enlarged to 80~mm ($\phi$80) for the UV laser in order to enable 
  the transformation of a laser beam shape as described in Section~\ref{sec:newtest2}. 
  In the measurements of beam intensities, a focal length was adjusted to increase 
  the laser-electron collision rate at best by changing the length between concave lenses 
  and a convex lens with a micrometer. 

 \subsection{Laser Injection Methods} \label{sec:inj}

    A laser beam enlarged at the beam expander was reflected at two high-reflection (HR) 
  coated mirrors, indicated as the third and fourth mirrors in Fig.~\ref{fig:hutch}, 
  before the injection into vacuum chambers. These two mirrors, whose diameters were 
  designed to be 80~mm, were mounted on a set of micro-stepping motors for a remote 
  control of horizontal rotation and vertical elevation angles. In this scheme, angle 
  variations and parallel shifts of a laser beam are possible for the precise adjustment
  of its axis relative to the electron beam. The LEP beam intensities with the injection
  of a single laser beam were basically measured by this mirror setup or putting 
  an additional mirror with a large effective area. A linear polarization vector of 
  a laser beam was controlled by a quartz half-wave ($\lambda/2$) plate, whose diameter 
  was 48~mm. It was placed just after the beam expander output in order to avoid a damage 
  by a high laser power density. A direction of the linear polarization was switched 
  vertically or horizontally by rotating the $\lambda/2$ plate. In the LEP intensity 
  measurements described below, the $\lambda/2$ plate was sometimes taken out from 
  the laser path.

    For the purpose of simultaneous injection with two UV lasers, we put two `Paladin'
  lasers in parallel as shown in Fig.~\ref{fig:hutch}. In the rear space of them, we 
  placed two `Frequad-HP' lasers on a higher stage for another simultaneous injection 
  of DUV laser beams. Two sets of the $\phi$80 beam expanders with manual linear and 
  rotation stages were installed in front of the `Paladin' lasers, while they were 
  replaced to the $\phi40$ beam expanders with higher stages in the case of using 
  the `Frequad-HP' lasers. Two laser beams, which were reflected at the third and 
  fourth mirrors, were incident on a right-angle prism from opposite sides, so that 
  both beam paths were directed to the storage ring side by side after the reflections 
  at two perpendicular planes with HR coating. Since the LEP intensity became higher 
  when a laser injection angle was close to the condition of a head-on collision 
  with the electron beam, the spacial position of the prism was adjusted by three 
  perpendicular linear stages. Laser beams were incident near the storage ring-side 
  edge on the reflection planes, whose width was 56.6~mm and 42.4~mm for the UV and 
  DUV laser injection, respectively. The prism size for the DUV laser injection was
  slightly smaller than the laser beam cross section, and this effect was taken into 
  account in Section~\ref{sec:2laser}.

    Alternative method of two-laser injection, which accomplishes complete head-on 
  collisions, may be possible by using a polarizing beamsplitter and overlapping 
  two beams with linear polarizations normal to each other. However, we chose 
  the method using a prism because the polarization vectors of two beams can be 
  aligned in the same direction as each other. The third and fourth mirrors and 
  the prism, all of which were made of synthetic fused silica, were separately 
  prepared for the UV and DUV lasers in order to optimize HR coating to 
  the individual wavelengths. We changed those optics alternatively depending on 
  the injected laser beams.

 \subsection{The LEPS Beamline} \label{sec:bl33lep}

    A laser beam was finally injected into vacuum chambers, connected to the storage 
  ring with the ultrahigh vacuum of $10^{-8}$~Pa. An entrance window of the vacuum 
  chamber is made of a 4~mm-thick synthetic fused silica, which is mounted on 
  an ICF-114 flange and possesses the effective area with a 60~mm diameter. 
  The injected laser beam was further reflected twice at aluminum coated silicon 
  mirrors toward the straight section of the storage ring. These mirrors are called 
  the first and second mirrors as shown in Fig.~\ref{fig:hutch}. Thickness of the 
  second mirror was designed to be~19 mm for ensuring surface accuracy, while that 
  of the first mirror was reduced to 6~mm because a LEP beam passes through this 
  mirror. Since X-rays due to synchrotron radiation are irradiated onto the first 
  mirror, it was cooled by a water flow from a rear holder, which has a hole for 
  the LEP beam path. 

\begin{figure}[htbp]
   \centering
   \includegraphics[width=8cm]{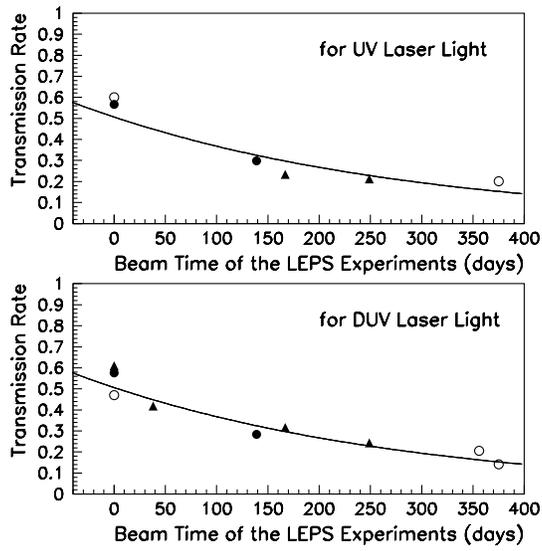}
   \caption{Total transmission rates from the entrance window to the exit of mirror 
            monitor chamber for UV (upper panel) and DUV (lower panel) laser beams. 
            The transmission rates are plotted depending on the beam time of the LEPS 
            experiments. The day when the entrance window and the first and second 
            mirrors are replaced to new ones is defined as zero in the horizontal axes. 
            Different symbols indicate different series of transmission measurements. 
            The absolute values of transmission rates are affected by the common 
            reduction at the monitor mirror and view port to extract a laser beam. 
            The fitting result expressed by Eq.~\ref{eqn:trans} is drawn by a solid 
            line in both panels. The fit was done by assuming a power measurement 
            error of 10\%.}
   \label{fig:trans}
  \end{figure}

    During a long-term use of these two mirrors and the entrance 
  window under a high radiation environment, we observed a drop of reflection and 
  transmission rates for both UV and DUV laser light due to the attachment of 
  hydrocarbons from a residual gas inside the chambers. Figure~\ref{fig:trans} 
  shows the change of a total transmission rate for the propagation through those 
  optical components as a function of beam time (or total time under radiation 
  exposure) at the LEPS experiments. The transmission rate was measured by taking 
  a ratio of laser powers before injecting the laser light to the entrance window 
  and after extracting it from the monitor mirror chamber, indicated in 
  Fig.~\ref{fig:hutch}. From a fit for all the data points in the upper and lower 
  panels of Fig.~\ref{fig:trans}, we have found this transmission rate ($T$) changes 
  by following:
  \begin{equation}
     T = exp{(-x/312.88)}, \label{eqn:trans}
  \end{equation}
  where $x$ denotes the beam time in the unit of days. It turns out that decreasing 
  rates of the transmissions for UV and DUV wavelengths are similar to each other. 
  Due to the observed deterioration, the entrance window and the first and second 
  mirrors have been periodically replaced in each 1--2 years. 

\begin{figure}[htbp]
   \centering
   \includegraphics[width=14cm]{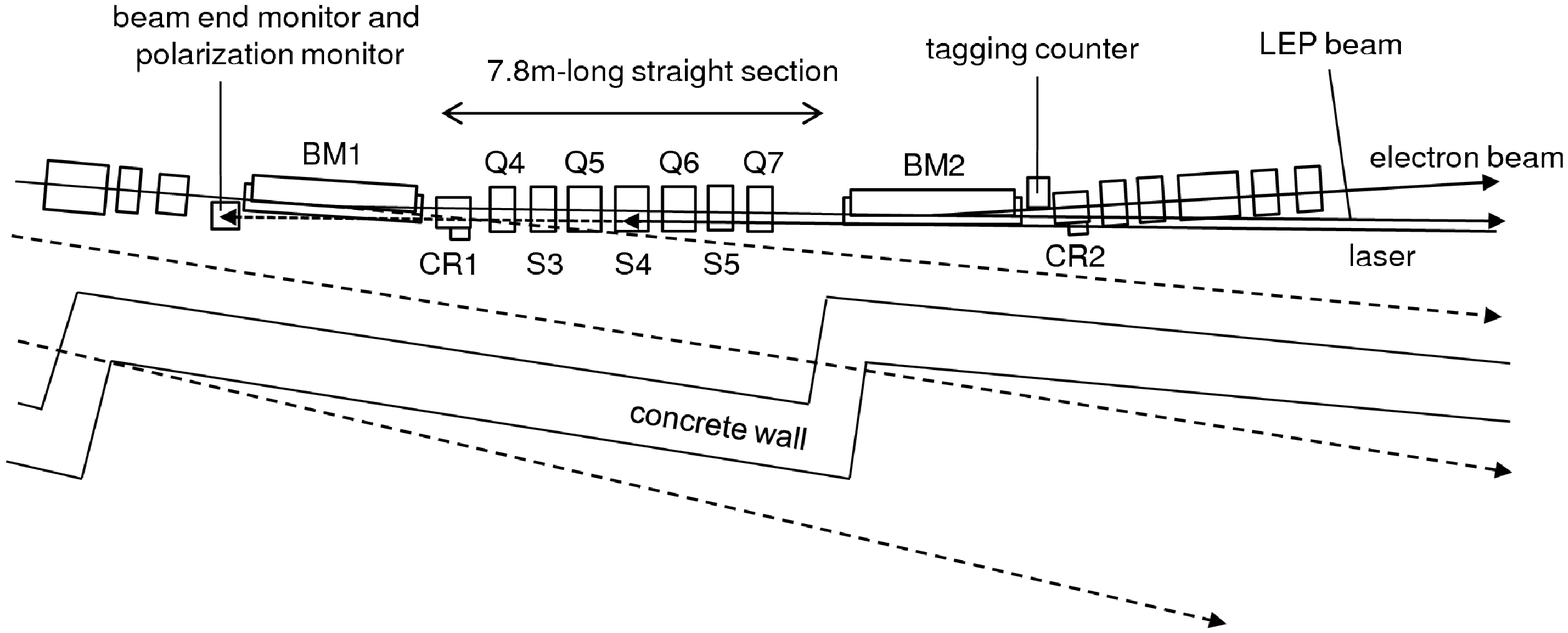}
   \caption{A plan view of the LEPS beamline around the 7.8~m-long straight section. 
            The `BM1--2', `Q4--7', `S3--5', and `CR1--2' represent bending magnets, 
            quadrupole magnets, sextupole magnets, and crotch absorbers, respectively.}
   \label{fig:bmend}
  \end{figure}

    The laser beam reaches the 7.8~m-long straight section, whose center is 
  the backscattering point. Figure~\ref{fig:bmend} shows a plan view of the LEPS 
  beamline around the straight section. An appropriate laser path was searched 
  for by adjusting the third and fourth mirror angles and confirming the laser 
  spot position at three monitor ports, where an insertion mirror or screen was 
  prepared. In Fig.~\ref{fig:bmend}, one of such monitor ports is indicated as 
  the `beam end monitor'. The best path was then achieved by maximizing 
  a laser-electron collision rate with the optimizations of mirror angles and 
  a prism position. Since the apertures of beamline chambers are limited due to 
  X-ray masks and magnets, the laser light only within a 40~mm $\times$ 40~mm 
  square region at the entrance window can pass to the collision point. While 
  the laser beam path in single-laser injection is not affected by the aperture 
  size, the propagation of two adjacent laser beams injected simultaneously is
  prevented by the limit of 20~mm on the horizontal width per beam. The width 
  of 20~mm corresponds to $\pm$1.4$\sigma$ and $\pm$1$\sigma$ in the horizontal 
  distributions of laser powers when using the `Paladin' and the `Frequad-HP', 
  respectively. This aperture effect was taken into account in Section~\ref{sec:2laser}.

    A backscattered photon beam, whose energy is magnified up to a GeV scale, 
  travels the ultrahigh vacuum beamline in the direction opposite to the laser 
  beams. After passing through the first mirror, the photon beam is extracted 
  from a 0.5~mm-thick aluminum window of the vacuum chamber inside the laser 
  hutch. (See Fig.~\ref{fig:hutch}.) X-rays in the beam are absorbed by 
  a 2.0~mm-thick lead plate, mounted on the aluminum window. The contamination 
  from e$^+$e$^-$ pair creations at the first mirror, the aluminum window, and 
  the lead absorber is removed by a sweep magnet, which provides a dipole magnetic 
  field of 0.6~Tesla. The cleaned-up photon beam is then transported from the 
  laser hutch to an experimental site, which is located about 70~m downstream 
  from the laser-electron collision point, through a medium vacuum pipe with 
  $\sim$1~Pa. A total transmission rate of the LEP beam from the collision point 
  to the experimental site was unavoidably reduced to about 60\% by 
  the pair-creations, mentioned above. More details of the LEPS beamline 
  facility are described in Ref.~\cite{muramatsu}.

 \subsection{Intensity Measurement} \label{sec:intensity}

    In the following sections, LEP beam intensities and energies were measured 
  by a tagging detector (or a `tagger'), which counts number of recoil electrons 
  in the backward Compton scattering and analyzes momenta of them event by event. 
  The tagger was installed at the electron beam exit of a bending magnet chamber, 
  as shown in Fig.~\ref{fig:bmend}. A recoil electron track is distinguished from 
  an electron beam orbit by a dipole magnetic field of 0.68~Tesla. Typically, 
  a deviation by 1~mm for the tagger hit position corresponds to a 30~MeV/c 
  difference of the recoil electron momentum. At early stages of the LEPS 
  experiments, the tagger was constructed from 10 fingers of trigger scintillators 
  and a 100~$\mu$m-pitch silicon-strip detector (SSD) with two layers. The LEP 
  energy spectrum discussed in Section~\ref{sec:duvinj} was measured by this 
  configuration. From the year 2005, the SSD part has been replaced to 1~mm-square 
  scintillation fibers with a 0.5~mm shift at different layers. The photon energy 
  was evaluated by subtracting the momentum of a recoil electron from the storage 
  ring energy of 7.975~GeV. A geometrical acceptance of the tagger restricts 
  the measurable range of the recoil electron momentum, so that the backscattered 
  photons are tagged at $E_{\gamma}$$>$1.5~GeV. Based on the integrations of the 
  backward Compton scattering spectra, the fractions of tagged LEP intensities 
  relative to the true intensities in the overall energy range are calculated to 
  be 38\% and 47\% for the injection of 355~nm UV and 266~nm DUV laser light, 
  respectively. A resolution of the measured LEP energy has been estimated to be 
  12~MeV, which is dominated not by a position resolution of the tagger but by 
  an electron beam divergence at the straight section.

    In the LEP intensity measurements by the tagger scintillators, dead time is 
  caused mainly by the rejection of accepting trigger hits during the time width 
  of a preceding signal. The dead time was originally 105~nsec for the tagger with 
  a SSD, but it was changed when the tagger was updated with scintillating fibers
  and the trigger logic was slightly modified. Most of the following intensity 
  measurements were performed with the dead time of 38~nsec. A correction factor 
  to be multiplied to the measured tagger rate was evaluated from the ratio of 
  hits rejected by the dead time in a random number simulation. This ratio depends 
  on the filling pattern of electrons, which are bunched inside the storage ring
  to 13~psec length by the 508.58~MHz RF frequency corresponding to the minimum 
  interval of 1.966~nsec. There are several variations for the filling pattern of 
  electron bunches, for example, by making a few hundreds of bunches isotropic on 
  the circumference of 1,436~m, or by filling several single bunches and a long 
  bunch train \cite{spring8}. An isotropic filling pattern generally lowers 
  the correction factor. The correction factor also varies depending on 
  the laser-electron collision rate. Therefore, it was formulated as a function 
  of the measured tagger rate for the individual filling pattern. While the correction 
  for dead time loss is applied in the following intensity measurements, we have not 
  converted the tagged LEP intensities to the values in the whole $E_{\gamma}$ range.

\section{Results of Photon Beam Upgrades} \label{sec:achievements}

 \subsection{Photon Beam Production using an All-Solid Mode-Lock UV Laser} \label{sec:modelock}

   The operation of an Ar laser (`Innova Sabre') needs a large electric power supply 
  and a huge cooling water flow because of large power consumption as described in 
  Section~\ref{sec:laser}. In addition, a periodical replacement of the argon-gas 
  tube was necessary to keep a reasonable output power. Hence, we have introduced 
  an all-solid mode-lock UV laser (8~W `Paladin'), whose electric power consumption 
  is significantly reduced and which requires only an internal water circulation 
  from an air-cooled chiller. The reduced power consumption also makes it possible 
  to operate multiple lasers simultaneously, as being discussed in Section~\ref{sec:2laser}. 
  A laser pointing at the 37~m-downstream focus and a UV output power of 8~W are 
  reasonably stable over several thousand hours, while the nonlinear THG crystal 
  must be occasionally shifted to avoid a damaged laser spot. Laser emission from 
  the `Paladin' is pulsed with a frequency of 80~MHz in a quasi-CW mode. Although 
  the electron beam is also bunched, a laser beam around the collision point possesses 
  a beam waist with a Rayleigh range of 0.8~m as shown in Fig.~\ref{fig:355waist}. 
  The 7.975~GeV electron encounters the 80~MHz laser photons at each 1.875~m, which is 
  roughly equivalent to twice the Rayleigh range. In addition, the minimum electron 
  bunch interval of 1.966~nsec, corresponding to a running distance of 0.6~m, is small 
  enough compared with the Rayleigh range. Therefore, this pulsed laser beam sufficiently 
  collides with electrons.

  \begin{table}[htbp]
  \caption{Measurements of LEP beam intensities for the injection of 7--8~W UV laser beams.
           There was a difference of the tagger dead time between the measurements with 
           the 8~W `Paladin' and the `Innova Sabre', resulting in the variation of correction 
           factors. The correction factor for the beam time difference is obtained so that 
           the LEP beam intensity should be scaled to the value at the radiation exposure 
           of 184~days by following Eq.~\ref{eqn:trans}.}
  \centering
  \begin{tabular}{lcccc}
  \hline
  Injected laser        & \multicolumn{2}{c}{8~W Paladin} & \multicolumn{2}{c}{Innova Sabre (7~W)} \\[5pt]
  \multirow{2}{*}{Polarization} &    Vertical    &    Horizontal   &    Vertical    &   Horizontal   \\
      & w/ $\lambda$/2 plate & w/ $\lambda$/2 plate & w/ $\lambda$/2 plate & w/ $\lambda$/2 plate \\
  \hline
  Tagger rate           &    1.15 MHz   &    0.99 MHz    &    0.84 MHz   &   0.99 MHz    \\[5pt]
  e$^-$ beam current    &    100 mA     &    100 mA      &    60.8 mA    &   83.5 mA     \\[5pt]
  \multirow{2}{*}{Filling pattern} & \multirow{2}{*}{203 bunch} & \multirow{2}{*}{203 bunch} & 203 bunch $-$      & 203 bunch $-$      \\
                        &           &           & 4 bunch $\times$ 7 & 4 bunch $\times$ 7 \\
  Tagger dead time      &    38~nsec    &    38~nsec     &    105~nsec   &   105~nsec    \\
  (Dead time corr.)     &($\times$1.043)&($\times$1.036) &($\times$1.107)&($\times$1.131)\\[5pt]
  Beam time             &    184 days   &    184 days    &    40 days    &   40 days     \\
  (Corr. factor)        &   (---)       &   (---)        &($\times$0.631)&($\times$0.631)\\
  \hline
  LEP intensity         & \multirow{2}{*}{1.20 MHz} & \multirow{2}{*}{1.03 MHz} & \multirow{2}{*}{0.96 MHz} & \multirow{2}{*}{0.85 MHz} \\
  for $E_\gamma > 1.5\;GeV$ &               &                 &                &                \\
  \hline
  \end{tabular}
  \label{tab:uv-intensity}
  \end{table}

    A LEP beam intensity with the 8~W `Paladin' was calculated from a tagger rate 
  and a dead time correction factor, as described in Section~\ref{sec:intensity}. 
  Table~\ref{tab:uv-intensity} summarizes the intensity calculations for the typical 
  experimental cycles covering 1--2 weeks. The tagger rate was measured in two cases, 
  where a linear polarization vector of the injected laser light was aligned in 
  the vertical or horizontal direction. For comparisons, corresponding numbers in 
  the case of using the `Innova Sabre' are also shown. Thanks to a top-up operation, 
  which keeps electron beam injection into the storage ring every one minute or less, 
  the tagger rates with the 8~W `Paladin' were always measured with an electron beam 
  current of 100~mA. In contrast, the measurements with the `Innova Sabre' were done 
  before the start of the top-up operation, so that the beam current of the storage 
  ring decreased with a life of a few tens to hundred hours depending on an electron 
  filling pattern. Thus, average currents of the electron beam during the intensity 
  measurements were evaluated for further corrections, as written in 
  Table~\ref{tab:uv-intensity}. All the tagger rates were measured in the periods when 
  203 electron bunches were isotropically filled in a circumference of the storage ring 
  or such a filling pattern was modified by removing 4 adjacent bunches from every 29 
  bunches. The difference of correction factors for the tagger dead time arises not 
  from the filling pattern but from the variation of measured tagger rate and 
  the change of dead time.

    In the case we used the 8~W `Paladin', the tagged LEP intensities 
  at the photon energies above 1.5~GeV were 1.20$\times$10$^6$~sec$^{-1}$ and 
  1.03$\times$10$^6$~sec$^{-1}$ for the vertical and horizontal polarizations, 
  respectively. The tagged beam intensity of $\sim$10$^6$~sec$^{-1}$ is well 
  reproduced by a naive model, where a `tube'-like flux of 355~nm laser light 
  ($\sigma=$150~$\mu$m) encounters a 100~mA electron beam ($\sigma_x=$295~$\mu$m
  and $\sigma_y=$12~$\mu$m) with the backscattering cross section of $\sim$0.20~barn
  at the interaction region defined by twice the Rayleigh range. By comparing
  the two measurements for different polarization states, we observed that the LEP 
  beam intensity with the injection of vertically polarized laser light was about 17\% 
  higher than that with the horizontally polarized laser injection. This is caused 
  by two horizontal reflections of laser light at the first and second mirrors, which
  are coated by aluminum. Based on the calculation with Fresnel coefficients at
  a 45$^\circ$ angle of incidence \cite{rakic}, a reflection rate of S-polarized UV
  light, whose electric field direction is perpendicular to a plane of incidence, 
  is $\sim$6\% higher than that of P-polarized one, whose electric field orientation 
  is parallel to the same plane. Thus, about 11\% difference is expected for the two
  successive reflections. This value is not far from the observed difference, and
  the actual measurements must be affected by surface oxidation and degradation.

    The LEP beam intensities with the `Innova Sabre' were measured to be 
  0.96$\times$10$^6$~sec$^{-1}$ and 0.85$\times$10$^6$~sec$^{-1}$ for the vertical and 
  horizontal polarizations, respectively. These intensities have been corrected by 
  taking into account the tagger dead time and the normalization at the electron beam 
  current of 100~mA. In addition, the difference of a laser transmission rate relative
  to the case with the `Paladin' has been cancelled by scaling the intensities at 
  the beam time of 184 days. Table~\ref{tab:uv-intensity} suggests that the LEP beam 
  intensity has been actually raised up after introducing the 8~W `Paladin'. 
  This improvement was achieved by a slight increase of the laser power and better 
  optimizations of laser injection optics. It is also important that the top-up 
  operation of the storage ring and the long-term stability of the laser power have 
  largely contributed to the net increase of LEP intensity.

 \subsection{Extension of the Maximum Photon Energy using DUV Lasers} \label{sec:duvinj}

\begin{figure}[htbp]
   \centering
   \includegraphics[width=8cm]{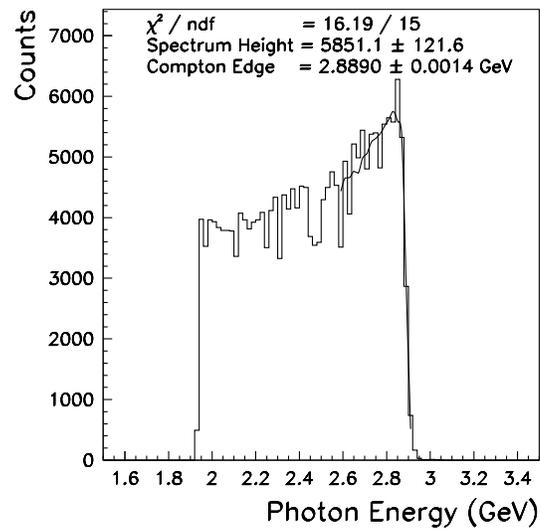}
   \caption{A LEP energy spectrum for the injection of 266~nm laser light. A template 
            spectrum is fitted with two parameters, which represent a spectrum height
            and a Compton edge.}
   \label{fig:3gev}
  \end{figure}

    The maximum energy of a LEP beam is extended by using a shorter wavelength laser, 
  as discussed in Section~\ref{sec:intro}. CW lasers with a power output of about 1~W 
  are currently available for the DUV wavelength of 266~nm or 257~nm, as described in 
  Section~\ref{sec:laser}. The generation of such a higher energy LEP beam was first 
  confirmed by operating a 266~nm laser (`DeltaTrain'). An energy spectrum of the LEP 
  beam was measured by the old tagger with a SSD, and the tagger energy calibration 
  was done by analyzing special data, where e$^+$e$^-$ pair production events from 
  a bremsstrahlung photon beam were collected by putting a 0.5~mm-thick lead converter
  plate at the LEPS experimental site. For the tagger calibration, momenta of the e$^+$ 
  and e$^-$ tracks were measured using the LEPS spectrometer \cite{muramatsu,leps-init} 
  with the particle identification by time-of-flight, and the measured momenta were 
  corrected by taking into account the energy loss due to bremssstrahlung radiation 
  at the lead plate and detector materials. The relation between the measured photon 
  energies, which were defined as a sum of the e$^+$ and e$^-$ momenta, and the tagger 
  SSD hit positions was formulated by fitting a second order polynomial function. A LEP 
  energy spectrum was then obtained by applying this energy calibration function to 
  a raw distribution of tagger SSD hit positions, which were recorded in the data
  accumulated by the tagger scintillator triggers with the 266~nm 
  laser injection. Here, good tagger hits were selected if a recoil electron track was 
  reconstructed by the two SSD layers and the trigger scintillators. Figure~\ref{fig:3gev} 
  shows the obtained spectrum after correcting the efficiencies of tagger scintillators. 
  The spectrum is largely fluctuated because of variations in SSD strip efficiencies. 
  The energy region lower than 1.9~GeV is not plotted because of a SSD module problem 
  in this measurement. By fitting a template spectrum from a backward Compton scattering 
  simulator, it was confirmed that the Compton edge was extended up to 2.89~GeV as expected.

    A typical DUV output power of the `DeltaTrain' was 0.7~W, and frequent manual 
  adjustments of inner optics were required. For physics data taking, a DUV laser based 
  on an argon-gas tube with a BBO crystal (`Innova Sabre Moto Fred') was instead introduced. 
  An output wavelength of this laser was 257.2~nm, so that the maximum energy of a LEP 
  beam slightly increased up to 2.96~GeV. Its output power was around 1~W, while there was
  variation depending on a seed laser power. The tagger rates with this laser were measured 
  for the two polarization states during a typical experimental cycle, as shown in the left 
  part of Table~\ref{tab:duv-intensity}. Tagger dead time corrections for an isotropic 
  filling of 203 electron bunches were estimated to be small because of low tagger rates. The 
  corrected LEP beam intensities at $E_\gamma$ $>$ 1.5~GeV were 0.21$\times$10$^6$~sec$^{-1}$ 
  and 0.16$\times$10$^6$~sec$^{-1}$ for the vertical and horizontal polarizations, respectively. 
  These LEP intensities are about one-tenth of those with the 8~W `Paladin', as recognized
  from the comparison of Tables~\ref{tab:uv-intensity} and \ref{tab:duv-intensity} together
  with taking into account about 40\% reduction of the UV light transmission due to the 
  additional radiation exposure of 161 days. The observed reduction by an order of magnitude
  can be understood by multiplying the ratios of laser powers (1~W / 8~W), laser photon 
  energies (257~nm / 355~nm), backward Compton scattering cross sections ($\sim$0.15~barn 
  / $\sim$0.20~barn), and tagged energy ranges (47\% / 38\%). The LEP beam whose energy was
  extended up to 3~GeV with the `Innova Sabre Moto Fred' was utilized for a few years, 
  resulting in wide physics outcomes as described in Ref.~\cite{hwang, fb20}.

  \begin{table}[htbp]
  \caption{Measurements of LEP beam intensities for the injection of 1~W DUV laser beams.
           The tagger rates with the `Innova Sabre Moto Fred' and the `Frequad-HP' were 
           measured after the radiation exposure of 23 days and 29 days, respectively.}
  \centering
  \begin{tabular}{lcccc}
  \hline
  Injected laser       & \multicolumn{2}{c}{Innova Sabre Moto Fred} & \multicolumn{2}{c}{Frequad-HP}  \\[5pt]
  \multirow{2}{*}{Polarization} &    Vertical    &    Horizontal   &    Vertical    &   Horizontal   \\
      & w/ $\lambda$/2 plate & w/ $\lambda$/2 plate & w/ $\lambda$/2 plate & w/ $\lambda$/2 plate \\
  \hline
  Tagger rate          &    0.21 MHz   &    0.16 MHz    &    0.25 MHz   &   0.21 MHz    \\[5pt]
  Filling pattern      &    203 bunch   &    203 bunch    &    203 bunch   &   203 bunch    \\
  (Dead time corr.)    &($\times$1.007) &($\times$1.005)  &($\times$1.009) &($\times$1.007) \\
  \hline
  LEP intensity & \multirow{2}{*}{0.21 MHz} & \multirow{2}{*}{0.16 MHz} & \multirow{2}{*}{0.25 MHz} & \multirow{2}{*}{0.22 MHz} \\
  for $E_\gamma > 1.5\;GeV$ &            &                 &                &                \\
  \hline
  \end{tabular}
  \label{tab:duv-intensity}
  \end{table}

    During the operation of the `Innova Sabre Moto Fred', the BBO crystal for SHG had to be 
  shifted frequently because the lifetime of a laser-irradiated spot was typically 1--2 weeks. 
  In addition, the argon-gas tube system as a seed laser caused huge power consumption and 
  required its periodical replacements. Therefore, we have changed it to a new solid-state 
  DUV laser (`Frequad-HP'), which provides CW emission with a wavelength of 266~nm. 
  The `Frequad-HP' has accomplished the low power consumption of 300~W only with air cooling 
  by adopting two stages of SHG from an IR diode laser. This feature enables simultaneous 
  two-laser injection as discussed in Section~\ref{sec:2laser}. A fixed output of 1~W stably 
  lasts over a few thousand hours because of a significant improvement of a BBO crystal 
  purity, which also increases a SHG efficiency up to $\sim$20\%. The right part of 
  Table~\ref{tab:duv-intensity} shows the results of LEP intensity measurements with 
  the `Frequad-HP'. For comparison with the case of the `Innova Sabre Moto Fred', 
  we examined tagger rates under similar conditions about the filling pattern and the beam 
  time. The tagged LEP beam intensities were evaluated to be 0.25$\times$10$^6$~sec$^{-1}$ 
  and 0.22$\times$10$^6$~sec$^{-1}$ for the vertical and horizontal polarizations, respectively. 
  These values are slightly larger than those with the `Innova Sabre Moto Fred'. There was 
  a gain by the stable DUV output. We had found that a narrow and high power beam from a DUV 
  laser seriously damaged the AR coating on the concave lens of a beam expander. However, 
  this damage was minimized enough after a power density was lowered by using the `Frequad-HP', 
  whose beam diameter was increased to 3.0~mm. This also contributed to obtaining a gain of 
  the LEP intensity during a long term data taking.

 \subsection{Increasing Beam Intensities by Two-Laser Injection} \label{sec:2laser}

    In principle, the flux of photons to be collided with electron beam is proportional 
  to the number of injected lasers. As already discussed, the simultaneous operation of 
  multiple lasers is now possible for both UV and DUV wavelengths at the LEPS experiments. 
  Table~\ref{tab:2laser} shows the results of LEP intensity measurements with the simultaneous 
  injection of two laser beams. The position of a prism, merging two laser beams, was scanned 
  in the three dimensional space in order to maximize tagger rates. At its optimum position, 
  the condition close to head-on collisions was satisfied, giving small injection angles 
  with moderate cutoffs of the expanded laser cross sections by the common edge of two 
  reflection planes. The measured tagger rate was corrected by the trigger dead time. 
  Since the tagger rates in this test were measured as short-time averages, the measurement
  error of them was about 10\%. In order to estimate the intensity gain, the tagger rate with 
  single-laser injection was also measured by rescanning the prism position, where a complete 
  head-on collision was achieved with the minimum cutoff of a laser cross section at 
  a reflection plane. This measurement was repeated twice for the two lasers, which were 
  used for the simultaneous injection. The two results were averaged in Table~\ref{tab:2laser}. 
  The LEP beam intensities with the simultaneous two-laser injection increased by factors of 
  1.5 and 1.6 in the cases of using the `Paladin' and the `Frequad-HP', respectively. High 
  statistics data collected with the two-laser injection is being analyzed for the studies 
  of exotic hadrons and so on \cite{fb20,theta06}.

    As explained in Section~\ref{sec:bl33lep}, the geometrical acceptance due to the beamline 
  aperture limits the LEP intensity gain of the two-laser injection. In order to estimate
  the acceptance-corrected gain factors expected at a larger aperture beamline, correction 
  factors were computed from the ratio of masked area in the laser cross section, as shown 
  in Table~\ref{tab:2laser}. For the case of single-laser injection, the correction for 
  the beamline aperture is not needed, but a finite prism size restricts the horizontal width 
  of laser beam paths as mentioned in Section~\ref{sec:inj}. Therefore, a small correction 
  factor due to this beam cutoff was obtained to estimate the true gain expected with a larger 
  prism, as shown in Table~\ref{tab:2laser}. The prism size correction was not taken into 
  account in the case of two-laser injection because that size was large enough compared 
  with the limitation from the beamline aperture. If all of these corrections are adopted,
  the acceptance-corrected gain factors of LEP beam intensities by the two-laser injection
  must reach 1.6 and 1.8 for the use of the `Paladin' and the `Frequad-HP', respectively. 
  These gain factors are close to two, while there is reduction by finite angles between 
  the electron beam and the laser beams around the collision point. 

  \begin{table}[htbp]
  \caption{LEP intensity gain factors by the simultaneous injection of two laser beams.
           The measurements with the `8 W Paladin' and the `Frequad-HP' were carried 
           out after the radiation exposure of 57 days and 75 days, respectively.
           Correction factors for the prism size were estimated based on the geometrical
           acceptance of reflection planes, which allowed the further propagation of
           355~nm and 266~nm laser beams only within the widths of 40~mm and 30~mm, 
           respectively.}
  \centering
  \begin{tabular}{lcc}
  \hline
  Injected laser                     & 8 W Paladin                      & Frequad-HP                      \\[5pt]
  \multirow{2}{*}{Polarization}      & Horizontal                       & Horizontal                      \\
                                     & w/o $\lambda$/2 plate            & w/ $\lambda$/2 plate            \\[5pt]
  Filling pattern                    & 203 bunch                        & 11 bunch train x 29             \\
  \hline
  2-laser injection                  & \hspace{1cm}                     & \hspace{1cm}                    \\
  \hspace{1cm} Tagger rate           & 3.0 MHz                          & 0.20 MHz                        \\
  \hspace{1cm} Dead time correction  & $\times$1.126                    & $\times$1.015                   \\
  \hspace{1cm} Aperture correction   & /0.956                           & /0.831                          \\[5pt]
  1-laser injection                  & \hspace{1cm}                     & \hspace{1cm}                    \\
  \hspace{1cm} Tagger rate           & 2.1 MHz                          & 0.12 MHz                        \\
  \hspace{1cm} Dead time correction  & $\times$1.082                    & $\times$1.009                   \\
  \hspace{1cm} Prism size correction & /1.000                           & /0.908                          \\
  \hline
  Intensity gain                     &                                  &                                 \\
  \hspace{0.5cm} only w/ dead time corr.           &      1.5           &     1.6                         \\
  \hspace{0.5cm} w/ all the corrections            &      1.6           &     1.8                         \\

  \hline
  \end{tabular}
  \label{tab:2laser}
  \end{table}

    In order to understand the observed gains, we performed a simple simulation, which 
  assumed the ideal propagation of a Gaussian beam. In this model, the power density of 
  a laser beam ($\rho$) at a radius of $x$ was calculated by:
  \begin{equation}
     \rho = \frac{2P}{\pi r^2} exp{(-\frac{2 x^2}{r^2})},
  \end{equation}
  where the 1/e$^2$ radius `$r$' varies along the beam axis as described in 
  Eq.~\ref{eqn:propagation}. The $P$ indicates the total power of a single laser. 
  The laser power densities in the three dimensional space overlapping with the electron 
  beam were numerically integrated using a Gaussian quadrature for the range of the BL33LEP 
  straight section, whose length is $\pm$3.9~m of the laser focus point. The electron beam 
  cross section, which is elliptical as described in Section~\ref{sec:focus}, was taken 
  into account by introducing the weight proportional to the electron density. While 
  the laser focal length was fixed at 37~m from the expander output, a horizontal position 
  of the expander was shifted by each 1~mm in order to scan injection angles. A part of 
  the laser cross section, corresponding to the area outside the reflection plane of 
  the prism in the adjacent beam side, was cut off at the numerical integration. The LEP 
  intensity gain by two-laser injection was calculated by doubling a result of the above 
  integration and then comparing it with a result from the single-laser injection with
  the injection angle of 0$^\circ$. This simulation was repeated for both wavelengths 
  of 355~nm and 266~nm by setting the 1/e$^2$ diameters of expander outputs to 28.6~mm 
  and 40~mm, respectively. The maximum gain factors in the injection angle scans were 
  obtained to be 1.48 and 1.52 for the 355~nm and 266~nm lasers, respectively. 
  The simulated results are comparable to the acceptance-corrected gain factors of 
  1.6 and 1.8 by taking into account the measurement error of tagger rates.

\section{Beam Intensity Upgrade for Near-Future Experiments} \label{sec:developments}

 \subsection{Test with a High Power Laser} \label{sec:newtest1}

    Increasing the laser power also leads to the proportional rise of a photon flux at 
  the collision point. There is no reduction of a LEP beam intensity due to a finite 
  injection angle in this case. Thanks to the progress in solid-state laser technologies, 
  the output power of a UV laser with the wavelength of 355~nm has recently reached 16~W
  by upgrading the 8~W version of `Paladin' laser. The injection of a 16~W laser beam 
  was tested at the LEPS beamline, resulting in the LEP beam intensity shown in 
  Table~\ref{tab:16w}. The uncertainty of the tagger rates was about 10\% because of 
  short-time measurements. In this test, a measurement with the 8~W laser was additionally
  performed using the same beam expander, which was designed for common use even with
  the different diameters of the 8~W and 16~W laser beams. By taking into account 
  the corrections for tagger dead time, it was confirmed that the LEP beam intensity 
  increased by a factor of $\sim$1.8 with the 16~W laser. This observed gain factor
  is close to the expected value of two.
  \begin{table}[htbp]
  \caption{Comparison of LEP beam intensities for the injection of 16~W and 8~W UV laser 
           beams. These measurements were done after the radiation exposure of 164--165 days.}
  \centering
  \begin{tabular}{lcccc}
  \hline
  Injected laser       & \multicolumn{2}{c}{16 W Paladin} & \multicolumn{2}{c}{8 W Paladin} \\[5pt]
  \multirow{2}{*}{Polarization} &    Vertical    &    Horizontal   &    Vertical    &   Horizontal   \\
      & w/ $\lambda$/2 plate & w/ $\lambda$/2 plate & w/ $\lambda$/2 plate & w/ $\lambda$/2 plate \\
  \hline
  Tagger rate          &    2.2 MHz     &    1.7 MHz      &    1.3 MHz     &   1.0 MHz      \\[5pt]
  Filling pattern      &    203 bunch   &    203 bunch    &    203 bunch   &   203 bunch    \\
  (Dead time corr.)    &($\times$1.087) &($\times$1.065)  &($\times$1.047) &($\times$1.037) \\
  \hline
  LEP intensity & \multirow{2}{*}{2.4 MHz} & \multirow{2}{*}{1.8 MHz} & \multirow{2}{*}{1.3 MHz} & \multirow{2}{*}{1.0 MHz} \\
  for $E_\gamma > 1.5\;GeV$ &            &                 &                &                \\
  \hline
  \end{tabular}
  \label{tab:16w}
  \end{table}

\begin{figure}[htbp]
   \centering
   \includegraphics[width=8cm]{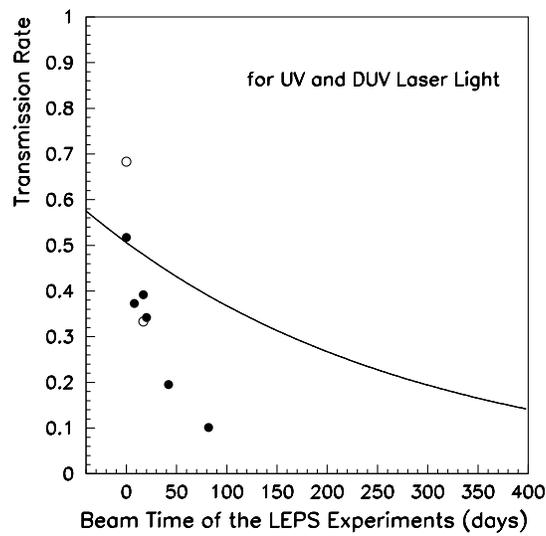}
   \caption{Total transmission rates from the entrance window to the exit of mirror monitor 
            chamber during the use of the 16~W UV laser. These values were measured before 
            improving a vacuum level. The transmission rates are plotted depending on 
            the beam time of the LEPS experiments. Closed and open circles indicate 
            the measurements with UV and DUV lasers, respectively. The solid line is the same 
            as the lines in Fig.~\ref{fig:trans}, and it is described by Eq.~\ref{eqn:trans}.}
   \label{fig:trans16w}
  \end{figure}

    During the 16~W laser tests, two problems were recognized for a long-term operation. 
  Firstly, a laser focus position transversely fluctuated more than a few hundred $\mu$m, 
  and it gradually drifted while requiring mirror angle adjustments every $\sim$10 minutes. 
  This problem is caused by the slight temperature variation at laser head optics. Therefore, 
  the cooling power of a chiller was raised up from 300~W to 1~kW. As a result, the laser 
  pointing stability was greatly improved, so that the tagger rate was retained without
  short-time drops. Secondly, serious damages were found at the first and second mirrors 
  and the entrance window after a long-term use of the 16~W laser. We observed the rapid 
  drop of a total transmission rate for them, as shown in Fig.\ref{fig:trans16w}. 
  By examining the damaged optical components after taking them out from the vacuum chambers, 
  we noticed more discoloration and lower transmission at the center parts of the second 
  mirror and the entrance window, which were placed under a lower vacuum condition 
  ($\sim$10$^{-7}$~Pa). It is deduced that the attachment of hydrocarbons from a residual 
  gas under X-ray radiation is serious with a high power UV laser. This was endorsed by 
  the detection of carbons on the damaged mirror surface as a result of Raman spectroscopy. 
  In order to improve a vacuum level by an order of magnitude, NEG pumps have been additionally 
  installed around the above optical components. The rapid drop of a total transmission rate 
  is now restored even with the 16~W laser.

 \subsection{Production of an Elliptical Laser Beam} \label{sec:newtest2}

    Another procedure we tested for the LEP intensity upgrade is the transformation of 
  a laser beam shape. A laser beam with a good emission mode (TEM$_{00}$) possesses 
  a round cross section, while the electron beam shape in the storage ring is horizontally 
  spread because of the synchrotron radiation. In order to increase their collision rate, 
  reshaping either of the beams should bring about more sufficient overlap of their cross 
  sections. Unfortunately, it is not realistic to produce a round electron beam because 
  a large space and huge costs are necessary to install many magnets. Instead, we chose 
  a method to generate an elliptical laser beam, which was rather easily handled by 
  adding optical components. In principle, a Gaussian beam propagates by following 
  Eq.~\ref{eqn:propagation}, where the increase of a laser diameter at a beam expander 
  output by a factor of two inversely reduces the diameter at a focus point by half.
  Therefore, we developed a small beam expander, which expands a beam cross section 
  twice only in one direction, by combining cylindrical concave and convex lenses. 
  Such a `cylindrical expander' was then mounted at the entrance side of the $\phi$80
  normal beam expander in order to generate a vertically shrunk shape at the beam
  waist. The laser power density is expected to be doubled at the focus point.

\begin{figure}[htbp]
   \centering
   \includegraphics[width=14cm]{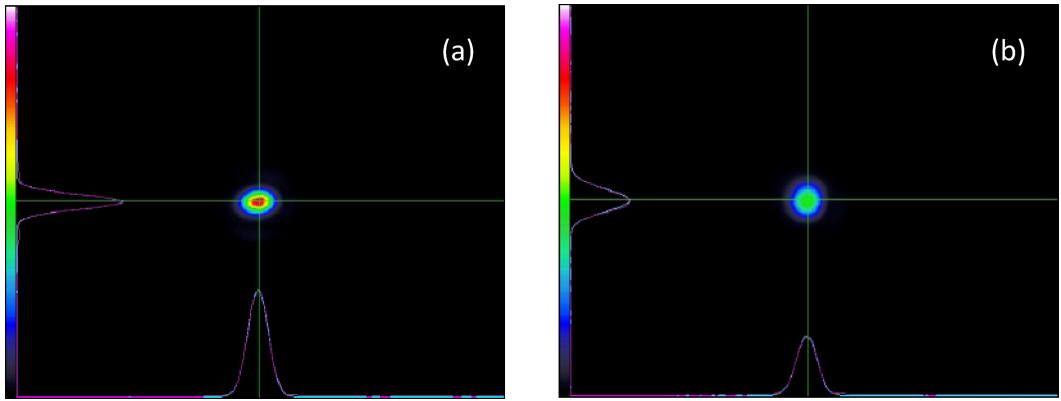}
   \caption{Laser beam profiles at the focus point (a) with and (b) without 
            a cylindrical expander. Twenty shots of the profile measurements 
            using the 8~W `Paladin' have been averaged at each panel.}
   \label{fig:bmprof}
  \end{figure}

    Before measuring the LEP beam intensity, we tested the cylindrical expander 
  by monitoring the beam shape with a beam profiler, made of a CMOS pixel sensor 
  (`LaserCam-HR-UV' produced by Coherent Inc.). Just because of space limitation, 
  the UV laser light from the 8~W `Paladin' was focused at $\sim$10~m downstream 
  of the beam expander by extracting the beam with extra mirrors and attenuating 
  the power with beamsplitters. Figure~\ref{fig:bmprof}(a) and (b) show the measured 
  profiles of laser powers with and without mounting the cylindrical expander on 
  the $\phi$80 expander, respectively. Gaussian fits to the vertical and horizontal 
  profiles, which are shown in the left and lower sides of the figures, resulted 
  in the 1/e$^2$ diameters as listed in Table~\ref{tab:bst}. The production of 
  an elliptical beam shape was confirmed with the reduction of a vertical diameter. 
  As shown in Table~\ref{tab:bst}, peak power densities were also measured by 
  the beam profiler in the two cases. By using the cylindrical expander, we observed 
  a gain in the power density by a factor of 1.81, which was close to the expected 
  value of two.

    From this result, the backscattering rate of laser photons must increase nearly 
  twice at the focus point. Simultaneously, the Rayleigh range or the effective range 
  for the laser-electron collisions becomes shorter by reducing the beam waist size.
  Therefore, we plan the next tests to extend the Rayleigh range with a slight reduction 
  of the peak power density by using a technique of a long range non-diffractive beam 
  (LRNB) \cite{lrnb}.

  \begin{table}[htbp]
  \caption{Peak power densities and 1/e$^2$ diameters of the 8~W UV laser beam 
           at a focus point with and without a cylindrical expander. In this
           measurement, the focal length was shorten to $\sim$10~m, and the laser
           power was attenuated by beamsplitters.}
  \centering
  \begin{tabular}{lcc}
  \hline
  \multirow{2}{*}{Setup}            & w/ cylindrical & only using $\phi$80 \\
                                    &    expander    &      expander       \\
  \hline
     Peak power density [W/cm$^2$]  &     1.072      &       0.591         \\
     Horizontal diameter [mm]       &     0.181      &       0.182         \\
     Vertical   diameter [mm]       &     0.127      &       0.194         \\
  \hline
  \end{tabular}
  \label{tab:bst}
  \end{table}

 \subsection{Simulation Results and Prospects for the New Beamline at SPring-8}

    Higher intensity LEP beams using the methods developed in the present article is 
  being produced at a new beamline of SPring-8 for the purpose to advance the next 
  generation hadron experiments, called LEPS2. The construction of the LEPS2 beamline 
  has started from 2010 at a 30~m-long straight section (BL31LEP), where a horizontal 
  divergence of the electron beam is significantly reduced to 12~$\mu$rad. Since 
  the produced LEP beam must be narrow enough even after a long propagation, a large 
  spectrometer system, covering $\sim$4$\pi$ solid angles with a 1~Tesla solenoidal 
  magnet, is under construction at a 135~m-downstream open space outside the storage 
  ring building. A large acceptance electromagnetic calorimeter system, called BGOegg, 
  is also prepared beside the charged spectrometer. At this new beamline, it is aimed 
  to increase the LEP intensity by about an order of magnitude for both the cases of 
  injecting UV and DUV laser light. Simultaneous injection of four laser beams are 
  intended by enlarging the apertures of beamline chambers.

    The four-laser injection is being achieved by preparing two sets of two-laser 
  injection systems, described in Sections~\ref{sec:inj} and \ref{sec:2laser}. 
  Two prisms, which individually merge two laser beams, are set up with a height 
  difference to avoid the interference of beam paths. We have designed to inject 
  the merged beams from a side-wall of the storage ring tunnel in order to reduce 
  the distance to a collision point. This way of laser injection minimizes 
  the apertures of beamline chambers, where an ultrahigh vacuum is necessary. 
  A radiation shield hutch surrounding the laser system is also needless because 
  the lasers are no more placed in the direction of synchrotron radiation. Four 
  `Paladin' lasers and four `Frequad-HP' lasers will be separately installed with 
  the injection optics, which are optimized for the individual wavelengths, inside 
  a clean room. Higher power lasers, including the newest type of `Paladin' with 
  an output power of 24~W, are also being introduced.

\begin{figure}[htbp]
   \centering
   \includegraphics[width=7cm]{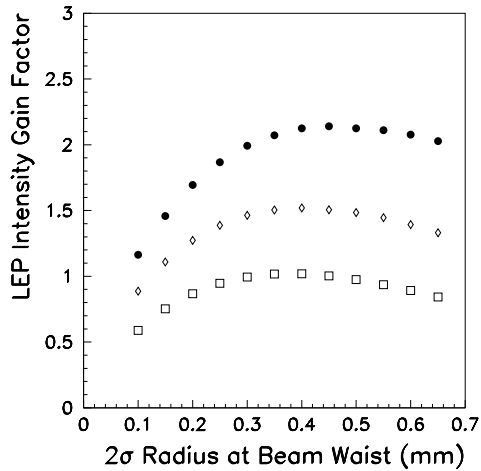}
   \caption{LEP intensity gains at the LEPS2 beamline relative to the single-laser 
            injection at the LEPS beamline. The cases of four-laser (closed circle), 
            two-laser (open diamonds), and single-laser (open squares) injection are 
            plotted as a function of the 2$\sigma$ radius of a beam waist. The use 
            of the 8~W `Paladin' was assumed in all the calculations.}
   \label{fig:4laser}
  \end{figure}

    Figure~\ref{fig:4laser} shows the estimation of LEP intensity gains by using 
  the 8~W `Paladin' lasers at the LEPS2 beamline in comparison with the case of 
  the single-laser injection using the same `Paladin' at the LEPS beamline. Closed 
  circles, open diamonds, and open squares indicate the relative gains depending on 
  the 2$\sigma$ (1/e$^2$) radius of a laser beam waist in the cases of four-, two-, 
  and single-laser injection, respectively. Here the ideal propagation of a Gaussian 
  beam was simulated in the way similar to that in Section~\ref{sec:2laser}. 
  A focal length of the laser light is reduced to 31.5~m at the LEPS2 beamline, and 
  the collision range at the straight section is increased to about 10~m, which is 
  determined from a constraint by the position of a laser monitor chamber. A transverse 
  position of the expander output center was scanned on a 1~mm matrix grid to search 
  for the best injection angle with the maximum gain, and the obtained maximum gain
  factors were plotted in Fig.~\ref{fig:4laser}.
  The area outside the reflection planes of prisms was cut off from the numerical 
  integration of power densities. The 1/e$^2$ radii of an electron beam at the long 
  straight section are 0.64~mm and 0.024~mm in the horizontal and vertical directions, 
  respectively, so that the LEP beam intensity by single-laser injection is maximized 
  at an average of these sizes. In contrast, the intensity by four-laser injection most 
  increases at a larger waist size because of a finite injection angle. The maximum 
  gain factor with four lasers is estimated to be 2.14.

    Although the simulation result is obtained by assuming the use of the 8~W `Paladin', 
  we actually operate the 16~W and 24~W types. We also plan to use the DUV lasers with 
  an increased power near future. Thus, an extra factor of two or more must be multiplied 
  to the estimated gain factor of 2.14. In addition, the optical handling of a laser cross 
  section is under considerations as discussed in Section~\ref{sec:newtest2}. Moreover, 
  an aluminum-coated silicon mirror, which guides the injected laser light to the straight 
  section, is modified to have a thin horizontal slit for avoiding X-ray irradiation and 
  a small hole for the LEP beam path. This reduces the total amount of e$^+$e$^-$ conversions
  during the LEP beam transportation to the experimental site, resulting in the further gain 
  of the transported beam intensity by a factor of 1.12. Such a mirror modification is
  possible only in the case of injecting four lasers.

    The construction of the LEPS2 beamline has just finished in the end of 2012. We have 
  observed the first LEP beam with single-laser injection in the beginning of 2013. 
  Optimizations for injection setups and intensity monitors have not been completed yet. 
  Detail descriptions and beam upgrade results for the LEPS2 beamline will be discussed 
  by a separate article. Experiments by the BGOegg detector will start in 2013, and 
  the spectrometer system is being prepared in parallel. Physics results with higher 
  precision and larger kinematical coverage, compared with the LEPS experiments, will 
  appear in the near future to obtain new insights of hadron natures.

\section{Summary} \label{sec:summary}

    At the SPring-8 LEPS beamline, we developed the methods to upgrade the high energy 
  LEP beam, which was produced by the backward Compton scattering of laser light from 
  7.975~GeV electrons. An all-solid UV laser with a wavelength of 355~nm was introduced 
  because of the advantage of low power consumption. In spite of the 80~MHz pulsed 
  emission, the output power of 8~W has successfully resulted in the LEP beam intensity 
  of 1--2$\times$10$^6$~sec$^{-1}$ in the tagged energy range of 1.5--2.4~GeV. There 
  is large variation of the measured intensity because the transmission rate at laser 
  injection optics drops with a life of 312.88 days by the accumulation of radiation 
  exposure. Thanks to the low power consumption, we performed the simultaneous injection 
  of two UV laser beams by merging them with a right-angle prism. We confirmed 
  an acceptance-corrected intensity gain by a factor of 1.6, which is reduced from two 
  due to small finite injection angles. The UV output power of the solid-state laser has 
  been improved up to 16~W or 24~W. We observed that the LEP beam intensity was nearly 
  doubled by using the 16~W laser.

    The maximum energy of a LEP beam has been also extended by introducing DUV lasers.
  We measured its energy spectrum by detecting recoil electrons at the tagger, and
  confirmed that the Compton edge was raised up to 2.89 GeV with a 266~nm laser as 
  expected. Using the solid-state laser with an output power of 1~W, the LEP beam
  intensity has reached 0.1--0.2$\times$10$^6$~sec$^{-1}$. Two DUV laser beams 
  were simultaneously injected in the way similar to the UV case, resulting in 
  the acceptance-corrected intensity gain by a factor of 1.8.

    The developed methods have been adopted at the LEPS beamline for hadron photoproduction 
  experiments. In addition, they are being utilized at the LEPS2 beamline, whose construction
  has just finished at SPring-8 in the end of 2012. As a result of simulations, four-laser 
  injection with the 24~W UV output powers increases the photon beam intensity at the LEPS2
  beamline by a factor of 6.4 relative to the case of the single 8~W laser injection at the
  LEPS beamline. An additional gain will be obtained from the transformation of laser beam 
  shapes. The tagged LEP intensity at 1.5$<E_\gamma<$2.4~GeV must be close to 10$^7$~sec$^{-1}$. 
  Similarly, we expect a higher intensity approaching 10$^6$~sec$^{-1}$ for the injection of 
  DUV laser beams. The laser mirror modification with a center hole for the LEP beam path 
  has a further effect to increase both the beam intensities at the experimental site by
  a factor of 1.12. Our developments would be applicable to other facilities to generate
  a laser-Compton backscattering beam.



\section*{}
  We thank to the support of the staff at SPring-8 for giving excellent experimental conditions.
  This research was supported in part by the Ministry of Education, Science, Sports and Culture 
  of Japan.

%% References


\bibliographystyle{elsarticle-num}
\bibliography{<your-bib-database>}

\begin{thebibliography}{00}


\bibitem{milburn}      R.H.~Milburn, Phys. Rev. Lett. 10 (1963) 75.
\bibitem{arutyunyan}   F.R.~Arutyunyan and V.A. Tumanian, Phys. Lett. 4 (1963) 176.
\bibitem{muramatsu}    N.~Muramatsu, arXiv:1201.4094 (2012).
\bibitem{dangelo}      A.~D'Angelo et al., Nucl. Instr. and Meth. A 455 (2000) 1.
\bibitem{ohgaki}       H.~Ohgaki et al., Nucl. Instr. and Meth. A 455 (2000) 54.
\bibitem{aoki}         K.~Aoki et al., Nucl. Instr. and Meth. A 516 (2004) 228.
\bibitem{kawase}       K.~Kawase et al., Nucl. Instr. and Meth. A 592 (2008) 154.
\bibitem{guo}          W.~Guo et al., Nucl. Instr. and Meth. A 578 (2007) 457.
\bibitem{ahn}          J.K.~Ahn and E.S.~Kim, Nucl. Instr. and Meth. A 528 (2008) 600.
\bibitem{spring8}      http://www.spring8.or.jp/.
\bibitem{leps-init}    T.~Nakano et al., Nucl. Phys. A684 (2001) 71c.
\bibitem{matsumura}    T.~Matsumura et al., Nucl. Instr. and Meth. A 582 (2007) 489.
\bibitem{hydroxyl}     W.P.~Leung et al., Appl. Phys. Lett. 58 (1991) 551.
\bibitem{rakic}        A.D.~Raki\'{c}, Appl. Opt. 34 (1995) 4755.
\bibitem{hwang}        S.H.~Hwang et al., Phys. Rev. Lett. 108 (2012) 092001.
\bibitem{fb20}         N.~Muramatsu, Few-Body Syst. (2013), DOI:10.1007/s00601-013-0614-4.
\bibitem{theta06}      Y.~Kato, Few-Body Syst. (2013), DOI:10.1007/s00601-013-0681-6.
\bibitem{lrnb}         T.~Aruga, Appl. Opt. 36 (1997) 3762.


\end{thebibliography}



\end{document}